\documentclass[11pt]{article}
\usepackage[margin=1in]{geometry}
\usepackage{hyperref}

\usepackage{cite}
\usepackage{xcolor}
\usepackage{amssymb, amsmath, amsfonts, amsthm}
\usepackage{mathtools}
\usepackage{algorithm}
\usepackage{multicol}
\usepackage{graphicx}
\usepackage{tikz}
\usepackage{array}
\usepackage{algpseudocode}
\usepackage{xspace}
\usepackage{subcaption}
\usepackage{booktabs}

\theoremstyle{plain}% default 
\newtheorem{theorem}{Theorem}[section] 
\newtheorem{lemma}[theorem]{Lemma}

\newtheorem{definition}{Definition}[section]

\theoremstyle{remark}

\DeclareMathOperator{\ratio}{ratio}

\DeclarePairedDelimiter\abs{\lvert}{\rvert}
\DeclarePairedDelimiter\norm{\lVert}{\rVert}

\makeatletter
\let\oldabs\abs
\def\abs{\@ifstar{\oldabs}{\oldabs*}}
\let\oldnorm\norm
\def\norm{\@ifstar{\oldnorm}{\oldnorm*}}
\makeatother

\newcommand{\sse}{\subseteq}
\newcommand{\ignore}[1]{}

\DeclareMathOperator{\EX}{\mathbb{E}}

\newcommand{\ECost}[1]{\EX[\text{cost}(#1)]}

\def\R{\mathbb{R}}
\newcommand{\opt}{\mathsf{OPT}}
\newcommand{\grd}{\mathsf{GRD}}

\newcommand{\alg}{\mathsf{ALG}}
\newcommand{\dks}{\ensuremath{{\sf DkS}}\xspace}
\DeclareMathOperator{\score}{score}

\DeclareMathOperator{\cov}{Cov}
\renewcommand{\emptyset}{\varnothing}
\newcommand{\batch}{\mathcal{B}}
\DeclareMathOperator{\poly}{poly}
\newcommand{\cvrp}{\ensuremath{{\sf CVRP_T}}\xspace}

\def\f{\ensuremath{{\cal F}}\xspace}
\def\sst{\ensuremath{{\sf SST}}\xspace}
\def\msc{\ensuremath{{\sf MSSC}}\xspace}
\def\qp{\ensuremath{{\sf QP}}\xspace}
\def\sbt{\ensuremath{{\sf SBT}}\xspace}
\newcommand{\reals}{\mathbb{R}}
\newcommand{\dre}{\ensuremath{{\sf DrE}}\xspace}
\DeclareMathOperator{\cost}{Cost}
\def\z{\mathbf{0}}

\begin{document}

\title{Sequential Testing with Subadditive Costs}
\author{Blake Harris\thanks{Department of Industrial and Operations Engineering, University of Michigan. Email: \{blakehar, viswa, rayent\}@umich.edu. Research supported in part by NSF grant CCF-2418495.} \and Viswanath Nagarajan$^*$ \and Rayen Tan$^*$}

\maketitle

\begin{abstract}
In the classic sequential testing problem, we are given a system with several components each of which fails with some independent probability. The goal is to identify whether or not some component has failed.
When the test costs are additive, it is well known that a greedy algorithm finds an optimal solution.
We consider a much more general setting with {\em subadditive} cost functions and provide a $(4\rho+\gamma)$-approximation algorithm, assuming a $\gamma$-approximate value oracle (that computes the cost of any subset) and a $\rho$-approximate ratio oracle (that finds a subset with minimum ratio of cost to failure probability).
While the natural greedy algorithm has a poor approximation ratio in the subadditive case, we show that a suitable truncation achieves the above guarantee.
Our analysis is based on a connection to the minimum sum set cover problem.
As applications, we obtain the first approximation algorithms for sequential testing under various cost-structures: $(5+\epsilon)$-approximation for tree-based costs, $9.5$-approximation for routing costs and $(4+\ln n)$ for machine activation costs. We complement these algorithmic results with some hardness results: (i)  sequential testing under submodular costs does not admit any poly-logarithmic approximation (assuming the exponential time hypothesis), and (ii) sequential testing under batch costs is NP-hard. 

\end{abstract}

\section{Introduction} \label{sec:introduction} 
Consider a manufacturing facility that needs to check a product for defects.
There are $n$ components in the product, each of which is defective with independent probability.
There is also a test for each component that incurs some cost: the test ``fails'' if the component is defective.
The goal is to identify whether any of the components has a defect.
Sequential testing involves performing tests one by one until either some defect is found or it is verified that there is no defect.
The objective is to minimize the {\em expected} total cost of tests performed.
It is well known that the sequence that orders tests in increasing order of cost to failure-probability is optimal~\cite{butterworth_reliability_1972,mitten_1960}.
In addition to the manufacturing application~\cite{duffuaa1990optimal}, sequential testing is applicable in healthcare~\cite{GreinerHJM06} and job-screening~\cite{Garey73} settings.

However, in many situations, the cost of testing is not additive (as in the setting above) because one might {\em batch} tests together and benefit from economies of scale.
Motivated by such considerations, there has been some recent work on sequential testing with {\em batch-costs}, where in addition to the individual testing costs (as above) there is a fixed setup cost $\beta$ that is incurred for every batch.
It has been observed that the batch-cost problem becomes much harder.
\cite{daldal2016approximation} obtained an approximation ratio $\approx 6.83$ and \cite{SegevS22} improved this to a PTAS.

While the batch-cost setting is one way of modeling economies of scale, it is still quite restrictive.
Our goal is to address more complex ``joint'' cost structures in sequential testing.
We mention two such examples here (see also Figure~\ref{fig:eg}).
\begin{itemize}
\item {\em Hierarchical (tree-based) costs.} The components in many systems have a modular structure, where they are arranged hierarchically in ``modules''.
In such cases, testing a component involves removing all modules that the component is contained in.
See \cite{Levi14} for an application in aircraft maintenance.
This cost structure can be modeled by a tree where each component is a leaf node and each module is an internal node.
The cost of testing a subset $S$ of components is the total cost of all nodes in the subtree induced by $S$.
The previously-studied batch-cost setting corresponds to a tree of depth one.
\item {\em Machine activation costs.} Suppose that the tests need to be performed on a set of machines, and each test has a specific subset of machines that can perform it.
If a machine is ``activated'' then it can perform all the tests that are allowed on it.
The cost of performing some subset of tests is the minimum number of machines that need to be activated for these tests.
\end{itemize}


\begin{figure}[t]
    \centering
    \resizebox{0.8\linewidth}{!}{\begin{tikzpicture}[x=1pt, y=1pt, line width=0.4pt,
    internal/.style={rectangle, fill, inner sep=0pt, minimum size=3.6pt},
    test/.style={circle, draw, fill=white, line width=0.6pt, inner sep=0pt, minimum size=3pt},
    title/.style={anchor=base}]

  \node[title] at (56,134) {(i) batch costs};
  \node[internal, label={[label distance=1pt]east:$\beta$}] (r) at (56,116) {};
  \node[test, label=below:$c_1$] (c1) at (16,84) {};
  \node[test, label=below:$c_2$] (c2) at (36,84) {};
  \node[test, label=below:$c_n$] (cn) at (96,84) {};
  \node at (66,84) {$\cdots$};
  \foreach \v in {c1, c2, cn} \draw (r) -- (\v);

  \node[title] at (190,134) {(ii) tree costs};
  \node[internal] at (190,116) {}
    [level distance=32pt,
     level 1/.style={sibling distance=48pt},
     level 2/.style={sibling distance=32pt},
     level 3/.style={sibling distance=16pt}]
    child {node[internal] {}
      child {node[test] {}}
      child {node[test] {}}}
    child {node[internal] {}
      child {node[internal] {}
        child {node[test] {}}
        child {node[test] {}}}
      child {node[internal] {}
        child {node[test] {}}
        child {node[test] {}}}}
    child {node[internal] {}
      child {node[test] {}}
      child {node[internal] {}
        child {node[test] {}}
        child {node[test] {}}
        child {node[test] {}}}};
  \node[title] at (198,8) {tests are leaf nodes};

  \node[title] at (350,134) {(iii) machine activation costs};
  \foreach \i/\y in {1/100, 2/68, 3/36} \node[internal] (m\i) at (326,\y) {};
  \foreach \i/\y in {1/116, 2/100, 3/84, 4/68, 5/52, 6/36} \node[test] (t\i) at (374,\y) {};
  \foreach \a/\b in {m1/t1, m1/t2, m1/t3, m2/t3, m2/t4, m2/t5, m3/t4, m3/t6} \draw (\a) -- (\b);
  \node[anchor=east] at (318,76) {machines};
  \node[anchor=west] at (382,76) {tests};
\end{tikzpicture}}
    \caption{Examples of subadditive cost structures.}
    \label{fig:eg}
\end{figure}

In this paper, we address such problems systematically by considering sequential testing problems where the cost-structure is a {\em subadditive} function.
We provide a generic approximation algorithm for subadditive sequential testing (\sst) where the approximation ratio depends on certain properties of the cost-function.
We also provide several applications of our result, which include tree-based and machine activation costs mentioned above.
We obtain the first approximation algorithms for these complex cost-structures.
We also show that some natural cost-structures cannot be approximated well.
In particular, submodular-cost sequential testing is harder to approximate than the densest-$k$-subgraph problem~\cite{Manurangsi17}, which essentially rules out any sub-polynomial approximation ratio (under the exponential time hypothesis).

\subsection{Problem Definition} \label{sec:problem-defn}
This paper deals with \textit{series systems}: systems that only function when all of its constituent components work.
There is a test associated with every component and the goal is to diagnose the system by identifying a defective component (if any).
Observe that it is not necessary to conduct all tests.
It suffices to find one component (test) that fails, in which case we know the system is defective, regardless of the outcome of other tests.
Testing is conducted until a failure is found (whereby we declare the system defective) or until all components are tested and deemed to work (whereby we declare the system working).
The objective is to minimize the expected cost of testing.

Formally, an instance of sequential batch testing consists of $n$ tests, given by $[n] \coloneqq \{1, \dots, n\}$.
Each test has a known independent probability $p_i$ of passing.
We let $q_i = 1-p_i$ be the probability of failure for each $i\in[n]$.\footnote{We assume that all the probabilities are provided as rational numbers and the instance size is $n\log L$ where $L$ is the largest integer in the rational representation.}
We represent the outcome of test $i$ as a random variable $X_i \in \{0,1\}$ where $X_i = 1$ if tests $i$ fails and $X_i = 0$ if test $i$ passes.

\subsubsection{Cost structure} 
We consider a very general cost structure that allows us to model several applications involving batched tests.
In particular, there is a family $\f\sse 2^{[n]}$ of subsets that represent allowed batches of tests.
Each batch $B\in \f$ also has a non-negative cost $c_B$.
This means that all tests in $B$ can be performed simultaneously at cost $c_B$.
Without loss of generality, the family \f is ``downward closed'', i.e., if $B\in \f$ and $B'\sse B$ then $B'\in\f$.
We note that the family \f and its costs may be specified implicitly: so $\abs{\f}$ may be exponentially large.
It will be convenient to work with a cost-function $c:2^{[n]}\rightarrow \R_+$ representing the cost to perform {\em any} subset $S$ of tests, i.e.,
$$c(S) = \min\left\{\sum_{B\in {\cal B}} c_B : S\sse \cup_{B\in {\cal B}} B \,,\, {\cal B} \sse \f\right\}, \quad \forall S\sse [n].$$

 The cost function $c(S)$ is the minimum cost of a collection of batches that ``covers'' $S$: so this is defined for every subset $S$. Moreover, the cost-function $c$ is monotone and subadditive.
Recall that function $c$ is \emph{subadditive} if $c(A)+c(B) \geq c(A \cup B)$ for all $A,B\sse [n]$.

Note that for any allowed batch $B\in \f$ we have $c(B) \le c_B$. Moreover, we can assume (without loss of generality) that $c(B)=c_B$ for all $B\in\f$. (If $c(B)<c_B$ then we just re-define the cost $c_B=c(B)$, which results in an equivalent instance.)  

A solution to the {\em subadditive series testing} (\sst) problem is given by a sequence $\batch = \langle B_1, \dots, B_k\rangle$ of batches that form a partition of $[n]$.
For any batch $B_j$ let $P(B_j)\coloneqq\prod_{i\in B_j} p_i$ denote the probability that all tests in $B_j$ pass.
Solution $\batch$ performs batch $B_j$ if and only if all tests in the preceding batches $B_1,\dots, B_{j-1}$
pass.
So, the probability that $B_j$ is performed is $\prod_{\ell=1}^{j-1} P(B_\ell)$ and the expected cost of the solution is:
$$\cost(\batch) = \sum_{j=1}^k \prod_{\ell=1}^{j-1} P(B_\ell) \cdot c(B_j).$$

Some special cases of \sst are listed below.
\begin{itemize}
\item In the classic series-testing problem, costs are additive, i.e., $c(S)= \sum_{i\in S}c_i$ where $c_i$ is the individual cost for test $i$.
\item In the batch series-testing problem, there is a setup cost $\beta$ in addition to the individual costs $\{c_i\}_{i=1}^n$ and $c(S)=\beta+\sum_{i\in S}c_i$.
\item In the tree-cost series-testing problem, there is a hierarchical structure on the tests given by a rooted tree with leaves corresponding to the $n$ tests.
Each node in the tree has a weight.
For any subset $S$, the cost $c(S)$ equals the total weight of all nodes in the subtree containing the leaves in $S$.
\end{itemize}

We will discuss more applications of \sst in \S\ref{sec:applications}.

\subsubsection{Oracles for the cost function} Our algorithm relies on two oracles for (approximately) solving certain optimization problems related to the cost function $c$.
The first oracle is a ``value'' oracle, which may itself be non-trivial as the cost structure is specified implicitly.
\begin{definition}[$\gamma$-approximate value oracle] \label{def:value-oracle} 
Given any subset $S\sse [n]$ this oracle returns a collection $\batch\sse \f$ of batches that covers $S$ with total cost $\sum_{B\in \batch}c_B\le \gamma\cdot c(S)$.
\end{definition}
In other words, we assume that there is a polynomial-time algorithm to obtain a $\gamma$-approximation to the cost of any subset.
The second oracle corresponds to the natural step in designing greedy algorithms for \sst.
\begin{definition}[$\rho$-approximate ratio oracle] \label{def:ratio-oracle} 
    Given any subset $U\sse [n]$ this oracle returns a batch $B^*$ that $\rho$-approximately minimizes the following ratio problem:
    \begin{equation}\label{eq:ratio}
        \min_{B\in\f, B\sse U} \,\quad \frac{c_B}{1-P(B)},
    \end{equation}
    where $P(B)=\prod_{i\in B} p_i$ is the probability that all tests in $B$ pass.
    In addition to the batch $B^*$, we assume that the oracle also returns an upper-bound ${C}^*$ on the cost $c_{B^*}$ such that $\frac{{C}^*}{1-P(B^*)}$ is at most $\rho$ times the minimum ratio in \eqref{eq:ratio}.
\end{definition}

Note that $1-P(B)$ is exactly the probability that testing will stop after batch $B$:
so this can be viewed as the ``benefit'' of selecting batch $B$, and the above ratio minimizes the cost-to-benefit ratio over all batches.
We could also have defined the ratio in \eqref{eq:ratio} by minimizing $\frac{c(B)}{1-P(B)}$ over {\em all} subsets $B\sse U$ (not just those in $\f$), which turns out to be equivalent.
Finally, the additional assumption that the ratio oracle returns an upper-bound $C^*$ on the cost is a mild assumption:
this holds in all our applications (and is a direct consequence of the $\rho$-approximation to the min-ratio value).

\subsection{Results and Techniques} \label{sec:results}
We present a generic approximation algorithm for \sst that relies on the approximate value and ratio oracles (Definitions~\ref{def:value-oracle} and \ref{def:ratio-oracle}).
\begin{theorem} \label{thm:main}
    There is a $(4\rho+\gamma)$-approximation algorithm for the subadditive series testing problem whenever there is a $\gamma$-approximate value oracle and a $\rho$-approximate ratio oracle.
\end{theorem}
Our algorithm uses the greedy approach that iteratively selects batches with the minimum cost-to-benefit ratio.
In fact, this ``ratio'' algorithm was already proposed in \cite{DaldalOSSU17} where it was evaluated computationally on a special case of \sst.
We first show that the approximation ratio of this simple greedy algorithm is $\Omega(\sqrt{n})$ even when $\gamma=\rho=1$.
In order to bypass such bad examples, we modify the greedy algorithm by considering all possible ``truncation'' points where instead of continuing with the greedy algorithm we just perform all remaining tests in a final batch.
Roughly speaking, our \sst solution is the truncated greedy solution that has the minimum expected cost.
In addition to the two oracles (Definitions~\ref{def:value-oracle} and \ref{def:ratio-oracle}), our analysis relies on a connection to the {\em minimum-sum set cover} problem~\cite{FeigeLT04}, for which a $4$-approximation algorithm is known.

Next, we show that several specific cost functions satisfy our assumptions, leading to good approximation algorithms for \sst under these cost structures.
Although the ratio problem~\eqref{eq:ratio} has a non-linear term in the denominator, we show that it can be linearized at the loss of a small approximation factor, which then allows us to use existing algorithms for ``quota'' versions of the respective covering problems.
Some of our applications are:
\begin{itemize}
    \item Hierarchical (tree) costs: we show that there is an FPTAS for the ratio oracle, which leads to a $(5+\epsilon)$-approximation algorithm for \sst.
    \item Routing costs: each test corresponds to a node in a metric and $c(S)$ is the minimum cost of a route that visits all nodes in $S$ from a root.
    We obtain a $9.5$-approximation algorithm for \sst using the TSP as the value oracle and the $k$-TSP problem~\cite{garg} as the ratio oracle.
    \item Machine activation costs: each test has an ``allowed'' subset of machines that may perform the test, and once a machine is activated it can perform all the tests that are allowed on it.
    $c(S)$ is the minimum number (or cost) of machines to ``activate'' to perform all tests in $S$.
    We obtain a $(4+\ln n)$-approximation algorithm for \sst using set-cover as the value oracle.
\end{itemize}

We also obtain some hardness results for \sst. 
\begin{itemize}
    \item 
First, we consider \sst when the cost function is {\em submodular}, which is a natural and well-structured class of subadditive functions.
Interestingly, this problem turns out to be very hard to approximate.
By establishing a relation to the dense-$k$-subgraph problem~\cite{Manurangsi17}, we prove that the submodular-cost series testing problem cannot be approximated to a factor better than $n^{1/\poly(\log\log n)}$ (assuming the exponential time hypothesis).\footnote{The exponential time hypothesis states that there is no $2^{o(n)}$ time algorithm for 3-SAT.}
The submodular function used in the hard instance is just a coverage function.

\item Next, we revisit the series testing problem under batch-costs and unconstrained batches, which is the simplest non-trivial special case of \sst. Here, the cost $c(S)=\beta + \sum_{i\in S} c_i$ for any $S\sse [n]$.  
As mentioned before, this problem has been studied in \cite{daldal2016approximation,SegevS22} and a PTAS was known. We show that series testing  under batch-costs is NP-hard, which resolves an open question from  these works.  
\end{itemize}

\subsection{Related Work} \label{sec:rel-work}
Sequential testing has been applied to many settings as mentioned earlier, and the greedy algorithm is known to be optimal for additive costs~\cite{butterworth_reliability_1972,mitten_1960}.
The extension to batched cost structures was proposed by \cite{DaldalOSSU17}, where the authors obtained some efficient heuristics (without approximation guarantees).
Later, \cite{daldal2016approximation} considered the batch-cost structure (with a setup cost) and obtained an approximation algorithm with ratio $\approx 6.28$.
This bound was subsequently improved to a PTAS by \cite{SegevS22}.
To the best of our knowledge, ours is the first paper to consider more complex cost-structures.

The minimum sum set cover problem (that is used in our analysis) is a central problem in approximation algorithms: it is well-known that a natural greedy algorithm is a $4$-approximation~\cite{Bar-NoyBHST98,FeigeLT04} and that this ratio cannot be improved unless $\text{P}=\text{NP}$.

Sequential testing problems have also been considered for other systems (beyond series systems).
Some examples are $k$-of-$n$ systems (which determines if at least $k$ components work)~\cite{ben1981optimal}, linear threshold functions~\cite{DeshpandeHK16} and score classification (which involves classifying the system based on the number of defects)~\cite{gkenosis_stochastic_2018,GhugeGN22}.
See also \cite{moret_decision_1982} and \cite{unluyurt_sequential_2004} for surveys.
All these results involve the simple additive cost structure.
A recent paper \cite{tanGeneralFrameworkSequential2024} obtains constant-factor approximation algorithms under batch-costs (with a fixed setup cost $\beta$ and unconstrained batches) for many of these systems.
However, this result relies heavily on this particular cost-structure and does not extend to the subadditive setting.

\subsection{Preliminaries on Minimum Sum Set Cover} \label{sec:mssc}
Our algorithm relies on a connection to the well-known {\em minimum sum set cover} (\msc) problem.
An instance of \msc consists of a set $E$ of elements with weights $\{w_e\}_{e\in E}$ and $M$ subsets $\{S_i\sse E\}_{i=1}^M$ with costs $\{c_i\}_{i=1}^M$.
An \msc solution is a permutation $\sigma=\langle \sigma(1), \sigma(2),\dots, \sigma(M) \rangle$ of the $M$ sets.
Given solution $\sigma$, the {\em cover-time} of any element $e\in E$, denoted $\cov(\sigma,e)$, is the cost of the smallest prefix of $\sigma$ that covers $e$.
That is, if $e\in S_{\sigma(j)} \setminus \left(S_{\sigma(1)} \cup \dots \cup S_{\sigma(j-1)} \right)$ then $\cov(\sigma,e) = c_{\sigma(1)} + \dots+ c_{\sigma(j)}.$
The objective in \msc is to minimize the total weighted cover time 
$$\sum_{e\in E} w_e\cdot \cov(\sigma,e).$$

The greedy algorithm for \msc works as follows.
If $R$ denotes the set of uncovered elements (initially $R=E$) then we select the set $S_i$ that minimizes the ratio $\frac{c_i}{\sum_{e\in S_i\cap R} w_e}$.

We will use the (known) result that this algorithm is a $4$-approximation~\cite{FeigeLT04}.
In fact, we need a more robust version that allows for approximate greedy choices, which also follows from prior work.

\section{Algorithm} \label{sec:alg}

The simple greedy algorithm iteratively chooses the batch $B\in \f$ (among the remaining tests) that minimizes the ratio $\frac{c_B}{1-P(B)}$.
This is known to be optimal in the special case of additive testing costs.
Moreover, this algorithm was proposed as a heuristic even for the batched setting in \cite{DaldalOSSU17}: there was no approximation bound known for it.
We first show that this greedy algorithm has approximation ratio $\Omega(\sqrt{n})$ for \sst.
Then, we present a modified greedy algorithm that truncates the greedy solution suitably, which leads to Theorem~\ref{thm:main}.

\paragraph{Bad Instance for Greedy.}
There are $n$ tests with cost function $c(S)=\min(\abs{S},\sqrt{n})$, which is \emph{subadditive}.
Each test $i\in [n]$ has failure probability $q_i\coloneqq\frac{1}{2^{i+1}}$.
Recall that $p_i=1-q_i$; so $p_1\le p_2\le \dots \leq p_n$.

We will show that the greedy solution performs tests in singleton batches, i.e., $\langle \{1\}, \{2\}, \dots, \{n\}\rangle$.
To this end, we show the following:
\begin{equation}\label{eq:bad-greedy} 
    \mbox{For any $j\ge 1$, the min-ratio batch in $\{j,j+1,\dots, n\}$ is $\{j\}$.} 
\end{equation}

To see this, let $U=\{j,j+1,\dots, n\}$ denote the remaining tests.
Fix any value $1\le k\le n-j+1$ and consider all batches $B\sse U$ with $\abs{B}=k$: it is clear that the minimum ratio among these batches is achieved by $B_k\coloneqq\{j,j+1, \dots, j+k-1\}$ because these are the tests with minimum $p_i$s.
We now have:
$$\frac{1}{2^{j+1}} = q_j \le \Pr[\mbox{some test in $B_k$ fails}] \le q_j+\dots +q_{j+k-1}< \frac{2}{2^{j+1}}.$$
So, the ratio of $B_k$ is 
$$\frac{c(B_k)}{1-P(B_k)} > 2^j \cdot \min(k,\sqrt{n}).$$
Moreover, the ratio of $B_1=\{j\}$ is exactly $2^{j+1}$.
It now follows that $\{j\}$ minimizes the ratio, proving \eqref{eq:bad-greedy}.

We now show that the expected cost of the greedy solution is at least $\frac{n}2$.
Indeed, the probability that {\em all} tests pass is at least $1-\sum_{i=1}^n q_i\ge \frac12$: under this event, the solution will have to incur cost of $\sum_{i=1}^n c(\{i\}) = n$.

\sloppy By performing all tests in a single batch, we get an expected cost of $\min\{n,\sqrt{n}\}=\sqrt{n}$.
So the optimal cost is at most $\sqrt{n}$.
This implies that the greedy algorithm has an approximation ratio of at least $\frac{n/2}{\sqrt{n}}=\frac{\sqrt{n}}{2}$.

\paragraph{Our Algorithm}
At a high level, the reason that greedy fails is that it does not utilize the property that testing {\em all} remaining components necessarily diagnoses the system (irrespective of the probabilities of failure).
This motivates our modified greedy algorithm that considers all possible ``truncation'' points, where all remaining tests are performed in one final batch.
The final solution is obtained by choosing the truncated greedy solution of minimum expected cost.

\begin{algorithm} \label{alg:mod-greedy}
\caption{Modified Greedy Algorithm}
\begin{algorithmic}[1]
\State set of remaining tests $U\gets[n]$; number of batches $\ell\gets0$ 
\While{$U \neq \emptyset$} \label{step:alg-greedy1}
    \State \label{step:alg-greedy-ratio} use the ratio-oracle to pick the batch $B_{\ell+1}$ that $\rho$-approximately minimizes
    $$\min_{B \subseteq U, B\in\f }\quad \frac{c(B)}{1-P(B)},$$
    along with an upper-bound $C_{\ell+1}$ on the cost $c(B_{\ell+1})$.
    \State $U\gets U\setminus B_{\ell+1}$ and $\ell\gets \ell +1$.
\EndWhile \label{step:alg-greedy2}
    
    \State \label{step:alg-grd-sol} let $\pi=\langle B_1,\dots, B_\ell\rangle$ be the greedy solution.
\For{$k=0, 1,\dots, \ell$}
    \State let $\pi_k$ be the truncated solution $\langle B_1,\dots, B_k, [n]\setminus \cup_{j=1}^k B_j \rangle$.
    \State use the value-oracle to obtain a $\gamma$-approximate value $D_k$ for batch $[n]\setminus \cup_{j=1}^k B_j$.
    \State define an upper-bound on the expected cost of solution $\pi_k$ as 
    $$G_k = \sum_{j=1}^k P(\cup_{h=1}^{j-1} B_h)\cdot C_j + P(\cup_{h=1}^{k} B_h)\cdot D_k.$$
\EndFor
\State Return the solution from $\{\pi_0,\pi_1,\dots, \pi_\ell\}$ with minimum upper-bound $\min_{k=0}^\ell G_k$.
\end{algorithmic}
\end{algorithm}

Observe that our algorithm relies on both the ratio oracle (Definition~\ref{def:ratio-oracle}) and the value oracle (Definition~\ref{def:value-oracle}).
We will show that this algorithm achieves a $(4\rho+\gamma)$ approximation, which would prove Theorem~\ref{thm:main}.
The analysis proceeds in two steps.
First, we bound the cost of the greedy solution $\pi=\langle B_1, B_2,\dots, B_\ell\rangle$ by relating it to an \msc instance.
Next, we show that there exists a truncation of the greedy solution that achieves a good approximation ratio.

\paragraph{Minimum sum set cover instance} 
Given any instance of \sst, we create an instance of \msc as follows: 
\begin{itemize}
    \item 
The elements are all non-zero vectors in $\{0,1\}^n$.
Each element $x\in \{0,1\}^n$ represents a \emph{realization} of all tests where test $i$ has outcome $x_i$.
The {\em weight} of any element $x$ is 
$$w_x = \prod_{i: x_i=1} q_i \cdot \prod_{i:x_i=0} (1-q_i)=\Pr[X_i=x_i,\ \forall i\in [n]],$$
the probability of realization $x$.
Note that we exclude the all-zero realization (where all tests pass).

\item There is a set $S_B$ for each batch of tests $B\sse [n]$ in \sst, where 
$$S_B= \{x\in \{0,1\}^n : \exists i\in B \, x_i=1\},$$ 
which are all realizations that have a ``fail'' outcome for some test in $B$.
The cost of set $S_B$ is just the cost $c(B)$ of the corresponding batch.
\end{itemize}

Although this \msc instance has an exponential number of elements and sets, it is not an issue because we only use this view in the analysis.

Recall that a solution to \msc is a permutation $\sigma$ 
of the sets, and $\cov(\sigma, x)$ denotes the cover time of any element $x$.
We note that the solution $\sigma$ need not be a full permutation: $\sigma$ can be any sequence of sets such that every element is covered.
The \msc objective is $\sum_{x\ne \z} w_x\cdot \cov(\sigma,x)$.
The greedy algorithm for \msc involves always selecting the set $S_B$ that minimizes the ratio:
$$\frac{c(B)}{\sum_{x\in S_B\cap R} w_x},$$
where $R$ is the current set of uncovered elements.
We will use the following result on the \msc algorithm with approximate greedy choices.
\begin{theorem}[\cite{FeigeLT04}]\label{thm:msc}
The \msc algorithm that always selects a set which is a $\rho$-approximation to the greedy criterion has approximation ratio $4\rho$.
\end{theorem}

For completeness, we provide a proof sketch in Appendix~\ref{app:mssc-proof}.

We now observe that the greedy sequence $\grd=\langle B_1, B_2,\dots, B_\ell\rangle$ constructed in Steps~\ref{step:alg-greedy1}-\ref{step:alg-greedy2} is exactly a $\rho$-approximate greedy solution to the above \msc instance.
Consider an iteration in the algorithm when $U\sse [n]$ is the set of remaining tests.
Irrespective of the exact batches chosen so far, the elements in the \msc instance that are still un-covered are 
$$R=\{x\in \{0,1\}^n \setminus \mathbf{0} : x_j=0 \; \forall j\in [n]\setminus U\}.$$
For any batch $B\sse U$, the new elements covered would then be:
$$S_B\cap R = \{x\in \{0,1\}^n \quad : \quad x_j=0 \; \forall j\in [n]\setminus U \mbox{ and } \exists i\in B : x_i=1\}$$
By definition of the weights $w_x$ and the fact that the outcomes $X_i$ are independent, we have
$$\sum_{x\in S_B\cap R} w_x = \Pr[X_j=0\; \forall j\in [n]\setminus U]\cdot \Pr[\exists i\in B : X_i=1] = P([n]\setminus U)\cdot (1-P(B)),$$
where we used the definition of $P(B)=\Pr[X_i=0\; \forall i\in B]$.
Therefore, the greedy criterion for \msc is:
$$\min_{B\sse U} \frac{c(B)}{P([n]\setminus U)\cdot (1-P(B))} = \frac{1}{P([n]\setminus U)} \cdot \min_{B\sse U} \frac{c(B)}{1-P(B)}, $$
where the equality is because the term $P([n]\setminus U)$ is a fixed value (not dependent on $B$).
So the greedy \msc criterion is just a scaled version of our greedy criterion in Step~\ref{step:alg-greedy-ratio}.
It follows that the greedy solution $\pi$ in Step~\ref{step:alg-grd-sol} is a $\rho$-approximate greedy solution to this \msc instance.

\paragraph{Relating \sst and \msc objectives.}
For any $x\in \{0,1\}^n$, let $\cov'(\pi,x)=\sum_{j=1}^k C_j$ where $B_k$ is the first batch that covers $x$.
(For $x=\z$ which is never covered, we set $\cov'(\pi,\z) = \sum_{j=1}^\ell C_j$.)
Note that $\cov'(\pi,x)$ is an upper-bound on the actual cover time $\cov(\pi,x)$ as each $C_j$ is an upper-bound on the cost of batch $B_j$.
Viewing the solution $\pi$ as a $\rho$-approximate greedy solution to \msc and using Theorem~\ref{thm:msc}, 
\begin{lemma} \label{lem:mssc-apx}
If $\pi^*$ is an optimal solution to the \sst instance then 
$$ \sum_{\substack{x \in \{0,1\}^n \\ x \neq \mathbf{0}}} w_x \cdot \cov'(\pi, x)\leq 4\rho \cdot \sum_{\substack{x \in \{0,1\}^n \\ x \neq \mathbf{0}}} w_x \cdot \cov(\pi^*, x).$$
\end{lemma}

The objective of any solution $\sigma$ to \sst can be written as 
\begin{equation}\label{eqn:expected_cost}
    \ECost{\sigma} = \sum_{\substack{x \in \{0,1\}^n \\ x \neq \mathbf{0}}} w_x\cdot \cov(\sigma, x) + w_{\mathbf{0}} \cdot \cov(\sigma, \z),
\end{equation}
where we explicitly separate the term for the all-zero realization $\z$, to highlight that \msc does not account for it.
Note that $\cov(\sigma, \z)$ is the maximum cost of running solution $\sigma$, which occurs when all tests pass.

\begin{figure}
    \centering
    \includegraphics[width=0.5\textwidth]{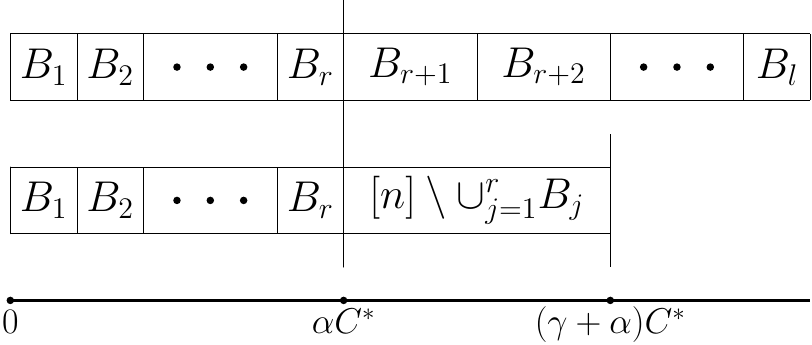}
    \caption{Ordering of greedy solution (Top) and the truncated solution (Bottom).}
    \label{fig:greedy-ordering}
\end{figure}

\paragraph{Truncating the greedy solution.} 
We define $C^* \coloneqq \cov(\pi^*, \z)$ to be the maximum cost in any run of the optimal solution $\pi^*$.
By subadditivity of the cost function, we have $C^*\ge c([n])$. Let $\alpha>0$ be some parameter that will be set later.
We now define a truncated greedy solution as follows: 
$$ \pi_r = \langle B_1, \dots, B_r, [n]\setminus \cup_{j=1}^r B_j\rangle,$$
where $r$ is the maximum index such that $\sum_{j=1}^r C_r \le \alpha C^*$.
See Figure~\ref{fig:greedy-ordering}.
We will show that the upper-bound on $\ECost{\pi_r}$ is $G_r\le (4\rho+\gamma)\cdot \ECost{\pi^*}$.
This would complete the proof of Theorem~\ref{thm:main}.

Below, we work with the cost upper-bounds $C_j$ for each batch $B_j$ (for $j\in [\ell]$) and $D_r$ for the last batch $[n]\setminus \cup_{j=1}^r B_j$ in $\pi_r$.
Recall that $\cov'(\pi,x)$ is the corresponding upper-bound on the cover-time of any $x\in \{0,1\}^n$ for the greedy solution $\pi$.
Similarly, we define $\cov'(\pi_r,x)$ to be the upper-bound on the cover-time of any $x\in \{0,1\}^n$ in the truncated solution $\pi_r$.

\begin{lemma} \label{lem:grd-mod-grd-rel}
    For each $x\in \{0,1\}^n$, $\cov'(\pi_r, x) \leq \left(1+\frac{\gamma}{\alpha}\right) \cdot \cov'(\pi, x).$
\end{lemma}
\begin{proof} 
    First, suppose that $\cov'(\pi,x) \le \alpha C^*$.
    Then we have $\cov'(\pi_r, x) = \cov'(\pi,x)$ as $x$ gets covered at the same point in both $\pi$ and $\pi_r$.

    Next, we claim that $D_r\le \gamma \cdot C^*$.
    Indeed, by monotonicity and subadditivity of the cost-function, we have $c([n]\setminus \cup_{j=1}^r B_j ) \le c([n])\le C^*$.
    So, using the $\gamma$-approximate value oracle, we get $D_r\le \gamma \cdot C^*$.
    
    Now suppose that $\cov'(\pi,x) > \alpha C^*$.
    We have 
    $$\cov'(\pi_r,x)\leq \cov'(\pi_r,\z) = \sum_{j=1}^r C_j + D_r \le \alpha C^* + \gamma C^* < \left(1+\frac{\gamma}{\alpha}\right) \, \cov'(\pi,x),$$
    which completes the proof.
\end{proof}

We are now ready to bound $G_r$.
\begin{align}
    G_r & = \sum_{\substack{x \in \{0,1\}^n \\ x \neq \mathbf{0}}} w_x\cdot \cov'(\pi_r, x) + w_{\mathbf{0}} \cdot \cov'(\pi_r, \z) \notag \\
    &\le \sum_{\substack{x \in \{0,1\}^n \\ x \neq \mathbf{0}}} w_x\cdot \cov'(\pi_r, x) + w_{\mathbf{0}} \cdot (\alpha+\gamma) C^* \label{eq:main-1}\\
    & \le \left(1+\frac{\gamma}{\alpha}\right) \sum_{\substack{x \in \{0,1\}^n \\ x \neq \mathbf{0}}} w_x\cdot \cov'(\pi, x) + w_{\mathbf{0}} \cdot (\alpha+\gamma) C^* \label{eq:main-2}\\
    & \le 4\rho \left(1+\frac{\gamma}{\alpha}\right) \sum_{\substack{x \in \{0,1\}^n \\ x \neq \mathbf{0}}} w_x\cdot
    \cov(\pi^*, x) + w_{\mathbf{0}} \cdot (\alpha+\gamma) C^* \label{eq:main-3}\\
    &= 4\rho \left(1+\frac{\gamma}{\alpha}\right) \sum_{\substack{x \in \{0,1\}^n \\ x \neq \mathbf{0}}} w_x\cdot
    \cov(\pi^*, x) + w_{\mathbf{0}} \cdot (\alpha+\gamma) \cov(\pi^*,\z)\label{eq:main-4} \\ 
    &\le \max\left\{\frac{4\rho}{\alpha} (\alpha+\gamma) \,,\, \alpha+\gamma \right\} \cdot \sum_{x \in \{0,1\}^n} w_x\cdot \cov(\pi^*, x) \notag\\
    &=\max\left\{\frac{4\rho}{\alpha} (\alpha+\gamma) \,,\, \alpha+\gamma \right\} \cdot\ECost{\pi^*} \notag
\end{align}
Above, \eqref{eq:main-1} uses the fact that $\cov'(\pi_r,\z) \le \alpha C^* + \gamma C^*$, \eqref{eq:main-2} uses Lemma~\ref{lem:grd-mod-grd-rel}, \eqref{eq:main-3} uses Lemma~\ref{lem:mssc-apx} and \eqref{eq:main-4} uses the definition of $C^*$.
Setting $\alpha=4\rho$, we obtain $G_r\le (4\rho+\gamma)\cdot \ECost{\pi^*}$ as claimed.

\subsection{The Ratio Oracle} \label{sec:greedy-choice} 
Here, we show how we can transform the ratio problem into a simpler constrained optimization problem known as the ``quota problem''.
This turns out to be very useful in our applications.
Recall that the ratio oracle requires finding the batch $B$ that minimizes $ \frac{c(B)}{1-P(B)}$.
Towards simplifying the ratio problem, we define $r_i = -\log p_i$ for all $i\in [n]$ and let $r(S)= \sum_{i\in S}r_i$.
Moreover, define the function $d(Q) = 1-e^{-Q}$, which is monotone and concave.
This allows us to write the ratio as $\ratio(B) = \frac{c(B)}{d(r(B))}$.
Using the above properties of function $d$, we can reduce the ratio problem to a linear-constrained optimization problem.

{\bf Minimum cost subject to quota (\qp).} Given a subadditive cost function $c:2^{[n]}\rightarrow\R_+$ and non-negative rewards $\{r_i\}_{i=1}^n$, and a quota $Q$ the goal is to find a subset $S\sse[n]$ of minimum cost such that the total reward is at least $Q$.
$$\min \left\{ c(S) : S\sse[n] \,,\,\sum_{i\in S}r_i\ge Q\right\}.$$

Compared to the ratio oracle (which has a non-linear term in the objective), the quota problem has a familiar knapsack-type structure found in optimization problems.
We rely on a bicriteria approximation for \qp:

\begin{definition}
\label{def:bicrit-qp}
    An $(\alpha, \beta)$-bicriteria approximation algorithm for \qp finds a subset $\hat{S}$ such that the cost $c(\hat{S}) \leq \alpha \cdot c(S^*)$ and $r(\hat{S}) \geq Q /\beta$, where $S^*$ denotes the optimal \qp solution.
\end{definition}

The next lemma shows how such an approximation for \qp can be used for the ratio oracle.
\begin{lemma} \label{lem: Quota}
    If there is an $(\alpha, \beta)$-bicriteria approximation for $\qp$, then there is a $((1+\epsilon) \alpha\beta)$-approximation for the ratio oracle for any fixed parameter $\epsilon > 0$.
\end{lemma}

Our proof uses the following observation:
\begin{lemma} \label{lem:d-ineq}
    Let $d:\reals \to \reals$ be function $d(x) = 1-e^{-x}$.
    For all $0<x\leq y$, we have $\frac{d(x)}{x} \geq \frac{d(y)}{y}$.
\end{lemma}
\begin{proof}
    The function $f(x) = d(x) / x$ has derivative $f'(x) = \frac{e^{-x}(x - e^x + 1)}{x^2}$.
    One can verify that the expression $x-e^x + 1\le 0$ for all $x\ge 0$.
    Therefore, $f'(x) \leq 0$ for all $x\ge 0$ and the result follows.
\end{proof}

We are now ready to prove Lemma~\ref{lem: Quota}.
\begin{proof}[Proof of Lemma~\ref{lem: Quota}]
The approximation algorithm relies on discretizing the search space of the quota $Q$ (representing sum of rewards) to obtain a polynomial-time approximation for the ratio problem.
We show how we can do so by incurring another $(1+\epsilon)$ loss for some $\epsilon > 0$.

We first set the boundary of the search space for $Q$.
The lowest non-zero value of $Q$ is $-\log p_{\max}$.
We bound $-\log p_{\max} = \log\frac{1}{1-q_{\min}} \geq q_{\min}$.
So, the lowest quota $Q$ is $r_{\min} = q_{\min}$.
The largest quota is set to be $r_{\max} = n \log \frac{1}{p_{\min}}$ which is an upper bound on the total reward.
Let $L = \mathcal{O}\left(\frac{1}{\varepsilon}\cdot \log \frac{n}{q_{\min}p_{\min}}\right)$.
For $i = 1, 2, \dots, L$, we set $\bar{Q}_i = r_{\min}(1+\varepsilon)^i$ and run the $(\alpha, \beta)$-bicriteria approximation for \qp with the quota $\bar{Q}_i$ to get the approximate solution $S_{i}$.
Finally, we return the subset $\hat{S} $ that minimizes the ratio:
$\min_{S_i} \frac{c(S_i)}{d(r(S_i))}$.

We claim that this gives us a $\mathcal{O}((1+\epsilon)\alpha\beta)$-approximation for the ratio oracle.
Moreover, the runtime is bounded by $\mathcal{O}\left(\frac{1}{\varepsilon}\cdot \log \frac{n}{q_{\min}p_{\min}}\right)$ calls to \qp, which is polynomial.

Based on our discretization, we know that:
$$r_{\min} (1+\varepsilon)^{i} \leq Q^*\leq r_{\min}(1+\varepsilon)^{i+1},$$
for some integer $i\geq 1$.
In addition, the value $\bar{Q}_i = r_{\min}(1+\varepsilon)^i \leq Q^*$ will be tried as the quota.
Note that $S_i$ is feasible to this \qp instance: so its optimal cost is at most $c(S^*)$.
By the bicriteria approximation guarantee, the corresponding solution $S_i$ has $r(S_i)\ge \bar{Q}_i /\beta\ge \frac{Q^*}{\beta(1+\epsilon)}$ and $c(S_i)\le \alpha\cdot c(S^*)$.
Applying Lemma~\ref{lem:d-ineq} yields $\frac{d(Q^*)}{d(r(S_i))}\leq \beta (1+\epsilon)$.
Hence,
$$\ratio(\hat{S}) \le \ratio(S_i)= \frac{c(S_i)}{d(r(S_i))} \leq \alpha\cdot\beta(1+\epsilon)\frac{ c(S^*)}{d(Q^*)} \leq \alpha\beta(1+\epsilon)\cdot\ratio(S^*),$$
which completes the proof.
\end{proof}

\section{Applications}\label{sec:applications}

In this section, we present applications of our results.
These instances are special cases of \sst that have good approximations for the value oracle and the ratio oracle.
In each of these cases, we summarize the approximation ratio for the ratio and value oracle.
In some cases, the ratio oracles require first finding an $(\alpha, \beta)$-bicriteria approximation for \qp (Defnition~\ref{def:bicrit-qp}), then deriving the ratio oracle using Lemma~\ref{lem: Quota}.
The approximation ratio for \sst is then obtained using Theorem~\ref{thm:main}.
A summary of our applications is presented in Table~\ref{tab:approx-table}.

\begin{table}
\centering
\caption{Approximation Ratio under different cost structure.}
\label{tab:approx-table}
\begin{tabular}{@{}l@{\hspace{5mm}}c@{\hspace{5mm}}c@{\hspace{5mm}}c@{}}
\toprule
\textbf{Cost Structure}             & $\rho$ & $\gamma$ & \textbf{Approximation Ratio} \\ \midrule
Concave cardinality                   &  $1$      &     $1$     &    $5$                 \\
Tree                       &    $1+\epsilon$    &     $1$     &      $5+\epsilon$               \\
Tree with batch capacity   &    $1+\epsilon$    &     $1+\epsilon$     &       $5+\epsilon$              \\
Machine Activation         &    $1$    &     $\ln n$     &          $4+\ln n$           \\
Routing (General Metric)   &     $2+\epsilon$   &   $1.5-\epsilon$       &           $9.5$          \\
Routing (Euclidean Metric) &   $1+\epsilon$     &  $1$        &            $5+\epsilon$         \\ \bottomrule
\end{tabular}
\end{table}

\subsection{Concave cardinality costs}
We consider settings where the cost of testing depends only on the {\em cardinality} of the batch (i.e., number of tests conducted), and the cost structure exhibits economies of scale.
Formally, there is a monotone increasing and concave function $g: \reals \to\reals$ such that the cost of testing any subset $S\sse [n]$   is given by $c(S) = g(\abs{S})$. 

For this problem,   we will show that the two oracles admit ratios $\gamma=\rho=1$.

\paragraph{Value Oracle.} As subset $S$ is a valid batch, computing the cost just involves evaluating the function $g$.
This sets $\gamma = 1$.

\paragraph{Ratio Oracle.} We can solve the ratio oracle optimally by renumbering the tests by increasing probabilities   $p_1 \leq p_2 \leq \dots \leq p_n$.
The algorithm then picks  the prefix $B_i = \{1, 2, \dots, i\}$ that minimizes the ratio $\frac{c(B_i)}{1-P(B_i)}$.
Enumerating all $i = \{1, 2, \dots, n\}$ only takes $\mathcal{O}(n)$ time.
It is easy to see that the optimal batch is among the prefix set $\{B_i\}_{i=1}^n$.
Indeed, as all batches of cardinality $i$ have the same cost $g(i)$, testing the size $i$ batch with the minimum $p_i$s minimizes the ratio.

By Theorem~\ref{thm:main}, this gives us a $5$-approximation.

It is worth noting that the bad instance for the natural greedy algorithm (in \S\ref{sec:alg}) involves concave cardinality costs. Our modified greedy algorithm achieves a constant approximation ratio in this setting: so   it    overcomes the limitation  of the natural greedy algorithm. We note however that \sst with  concave cardinality  costs can  be solved exactly with a dynamic programming algorithm.
The dynamic program relies on the symmetry in the cost structure, which implies that the batched solution is an ordered partition of the tests sorted by increasing probabilities $p_i$.

\subsection{Hierarchical (tree) cost structure} \label{app:tree-cost}
An instance of \sst with a hierarchical cost structure consists additionally of a weighted tree $T = (V, E)$ rooted at $r$.
Each of the $n$ tests is a leaf-node, where test $i$ is placed on leaf $v_i$.
Moreover, every node $v\in V$ is assigned a cost $w_v$, and can be \emph{activated} by incurring cost $w_v$.
Conducting test $i$ requires every node along the $r-v_i$ path to be activated.
Intuitively, subadditivity of this cost structure comes from the idea that two leaves $v_i$ and $v_j$ may share a common ancestor $a$, so nodes along path $r-a$ are charged only once.
Without loss of generality, by adding zero-cost nodes, we assume that $T$ is a binary tree.

\paragraph{Value Oracle.}
To calculate the cost of performing any subset $S$ of tests, we just need to add the weights of all nodes in the subtree induced by $S$.
So $\gamma = 1$.

\paragraph{Ratio Oracle.} 
We provide an approximation algorithm for \qp.
We assume without loss of generality that the root has zero reward (otherwise, we can add another 0-reward node above the root).
With this assumption, we can work with a tree cost structure where the \emph{edges} are activated.
For each edge $uv$, let $w_{uv} = w_v$, where $v$ is the child of $u$.
Conducting a test $i$ now requires a path of active edges from $r-v_i$.

We derive a PTAS for \qp using a dynamic program (DP).
The DP maximizes reward subject to a budget constraint, so the solution to \qp has to be recovered by guessing the lowest budget $b$ such that the reward is at least $Q$.
Each stage of the DP solves the reward maximization problem on some subtree of $T$ rooted at $u$, which we denote as $T_u$.
The state $b$ tracks the budget allocated to edges in $T_u$.
At stage $u$, the DP chooses whether to include and how much budget to allocate to its left and right subtrees.
The base case is defined on the leaves: $P(l, b) = r_l$ for every $b$.
Letting $v_1, v_2$ be the left and right child of $u$, the recurrence is given by: 
\begin{equation} \label{eq:Tree-DP}
P(u, b) = \max\left.
  \begin{dcases}
    &0,\\
    \qquad\max_{\substack{b_1 \in \R:\\ w_{uv_1}+b_1 \leq b}} & P(u_1,b_1),\\
    \qquad\max_{\substack{b_2 \in \R:\\ w_{uv_2}+b_2 \leq b}}& P(u_2,b_2),\\
    \max_{\substack{b_1, b_2 \in \R\\w_{uv_1}+w_{uv_2}+b_1+b_2 \leq b}} & P(u_1,b_1)+P(u_2,b_2),
  \end{dcases}
  \right\}.
\end{equation}

To achieve a polynomial time algorithm, we follow the standard rounding procedure for the knapsack problem.
That is, we fix some error parameter $\epsilon > 0$, and set $\mu = \epsilon B / n$ where $B$ is the budget allocated at the root.
We then round down all costs to integer multiples of $\mu$.
This gives us a $(1+\epsilon)$-approximation for \qp for any $\epsilon > 0$, and by Lemma~\ref{lem: Quota} we get a $(1+\epsilon)$-approximation for the ratio problem.

Applying the above value and ratio oracles gives us a $(5+\epsilon)$-approximation by Theorem~\ref{thm:main}.

\subsection{Tree Cost-Structure with Batch Capacity}

We consider the setting of \S\ref{app:tree-cost} with the addition of a new constraint that tests can only be conducted in batches of size up to $k$.
i.e., the family of allowed batches $\f = \{B: B \subseteq [n],\ \abs{B} \leq k\}$.

\paragraph{Ratio Oracle.} This can be solved by a simple extension of the dynamic program in \S\ref{app:tree-cost}.
In particular, the state of the DP can be augmented with another variable $h$ that denotes the maximum number of tests selected from the subtree: $P(u,b,h)$ is the maximum reward obtained from subtree $T_u$ such that the cost of edges is at most $b$ and the number of selected tests is at most $h$.
So, we again obtain a PTAS for \qp and the ratio oracle.

\paragraph{Value Oracle.} The value oracle here turns out to be non-trivial.
In fact, it can be viewed as an instance of \emph{Capacitated Vehicle Routing Problem on Trees} (\cvrp).
Given a subset of tests $S\sse [n]$, let the corresponding leaves $\{v_i\}_{i\in S}$ be ``demand points'' with unit demand, and let the demand of every other node be 0.
Let the root $r$ be the supply depot.
Moreover, set the distance of edge $uv$ as $d_{uv} = w_v / 2$, where $v$ is the child of node $u$.
An instance consists of a vehicle with capacity $k$, where the vehicle has to return to $r$ to refill once it delivers $k$ goods.
The goal of \cvrp is to route the vehicle such that the distance traveled is minimized.

Given any solution to \cvrp, each roundtrip from $r$ taken by the vehicle corresponds to a batch of tests $B$.
The capacity on the vehicle ensures that each batch is at most size $k$.
Moreover, let the set of edges traversed by the vehicle be ``active''.
On a tree, edges traversed by an $r-r$ tour will be visited exactly twice, so the distance traveled by the agent is exactly the cost of testing a batch $B$.
\cite{MathieuZ23} provides a PTAS for \cvrp, so we have $\gamma=1+\epsilon$ for any constant $\epsilon>0$.

This gives us a $(5+\epsilon)$-approximation by Theorem~\ref{thm:main} for every $\epsilon > 0$.

\subsection{Machine Activation Cost} \label{sec:or_testing}
In this setting, testing cost is defined on a set of $m$ machines.
Each machine $j\in [m]$ is capable of testing only some subset $T_j\sse [n]$ of tests, and has an activation cost $c_j$.
For each test $i$, we denote the subset of machines capable of performing test $i$ as $M_i = \{j \in [m]: i \in T_j\}$.
To perform test $i$, at least one machine $j\in M_i$ has to be activated.
Once a machine $j$ is activated, all the tests in $T_j$ can be performed.

\paragraph{Ratio Oracle.} 
Consider any arbitrary subset of tests $S\sse [n]$.
Suppose that testing $S$ requires turning on some subset of machines $K\sse [m]$ where $\abs{K} > 1$.
We can decompose the ratio into 
\begin{align*}
    \frac{c(S)}{1-\prod_{t\in S}p_t} &= \frac{\sum_{j \in K}c_j}{1-\prod_{t\in S} p_t} \geq \frac{\sum_{j\in K} c_j}{1-\prod_{t\in \bigcup_{j\in K} T_j}p_t}
    &\geq \frac{\sum_{j\in K} c_j}{\sum_{j\in K}\left(1-\prod_{t\in T_j} p_{t}\right)} \geq \min_{j\in K}\frac{c_j}{1-\prod_{t\in T_j}p_t},
\end{align*}
where first inequality follows from $S\sse \bigcup_{j\in K} T_j$, the second follows from submodularity of the set function $g(S) = 1-\prod_{t\in S} p_t$.
The last inequality follows from the mediant inequality.
Thus, the optimal batch is always a set $T_j$ for some $j\in [m]$.
So, the ratio problem can be solved in polynomial time by computing the ratio of $T_j$ for each $j\in [m]$ and picking the largest one.
Hence, $\rho = 1$.

\paragraph{Value Oracle.} Given some batch $B\sse [n]$ of tests to conduct, determining the optimal machines to activate is the classic {\em set cover} problem, which is NP-hard, but admits a $\ln n$ approximation algorithm.
So, $\gamma= \ln n$.

Then by Theorem~\ref{thm:main}, we have a $(4+\ln n)$-approximation.
We note that \sst under machine activation costs is at least as hard to approximate as set-cover.
Indeed, if the instance has all probabilities $p_i\rightarrow 1$ then \sst solutions are precisely set cover solutions.
Using the hardness of approximation for set-cover~\cite{DBLP:conf/stoc/DinurS14}, there is no approximation ratio better than $\ln n$ for \sst under machine activation costs.

\subsection{Routing Costs}
Here, tests are located at nodes of a metric space.
Testing a batch $B$ requires routing an agent from a root node $r$ to every test $t\in B$ and returning to $r$.
Let $(V\cup\{r\}, d)$ be a metric, where each vertex $v\in V$ is a test and $d:V\times V \to \reals_{\geq 0}$ is a distance function that satisfies symmetry and triangle inequality.

\paragraph{Ratio Oracle.}
This is again based on the quota problem.
We are given rewards $r_v$ on each vertex $v\in V$, and a quota $Q$.
We use the function $V(T)$ to denote the set of vertices visited along $T$, so $r(V(T)) = \sum_{v\in V(T)} r_v$ is the total reward collected along tour $T$.
The goal is to find the minimum cost tour $T$ such that $r(V(T)) \geq Q$.

Using simple scaling arguments, we show that the quota problem reduces to the well-known {\em $k$-Traveling Salesman Problem}~\cite{garg}.
We can assume, without loss of generality, that $r_i \leq Q$ for every $i$.
(If the optimal solution visited any vertex with reward more than $Q$, it is easy to find the solution by enumerating over all such vertices.)
Let $O^*$ be the optimal tour for the original \qp.
We define the scaling factor $r_0 = \frac{Q}{n^2}$ and round down each $r_i$ to some integer multiple of $r_0$.
That is, we set the rounded values $\bar{r}_i = \max_{z \in \mathbb{Z}_{\geq 0}}\{zr_0: zr_0 \leq r_i\}$.
Since we have $n$ nodes, and each node is rounded down by at most $r_0$, the reward of tour $O^*$ on this new instance is $\bar{r}(O^*) \geq r(O^*) - n\cdot \frac{Q}{n^2}=Q-\frac{Q}{n}\eqqcolon\bar{Q}$.
We now solve the instance with rewards $\bar{r}$ and quota $\bar{Q}$; by scaling all rewards by $r_0$ we obtain an instance where each reward is an integer value between $1$ and $n^2$.
Such an instance can be solved directly by $k$-TSP by replacing each vertex $i$ with $\bar{r}_i/r_0$ many co-located copies (note that the total number of vertex copies is polynomial).
Thus, we can use an existing $2$-approximation for $k$-TSP~\cite{garg} to obtain a $(2,1+\frac1n)$-bicriteria approximation to \qp.
Using Lemma~\ref{lem: Quota} we obtain a $(2+\epsilon)$-approximation for the ratio problem for any $\epsilon>0$.

\paragraph{Value Oracle.} This is just the classic TSP problem.
So, Christofides' algorithm gives $\gamma=1.5$.
We can also use the recent improvement to $1.5-\epsilon$ from~\cite{KarlinKG21}.

Using the above ratio and value oracles, applying Theorem~\ref{thm:main} gives a $9.5$-approximation for \sst under routing costs.
The approximation ratio can be further improved when the metric is  Euclidean: in this case, there is a PTAS for $k$-TSP~\cite{Arora98}, so we obtain a $(5+\epsilon)$-approximation for any constant $\epsilon > 0$.

\section{Hardness Results} 
\subsection{Approximation Hardness for \sst with Submodular Costs}
In this section, we prove a hardness of approximation for sequential testing with submodular costs (which is a natural special case of subadditive costs).
\begin{theorem}\label{thm:submod-hardness}
    \sloppy Assuming the exponential time hypothesis, there is no $n^{1/\poly(\log \log n)}$-approximation algorithm for submodular-cost sequential testing.
\end{theorem}
In particular, this rules out any poly-logarithmic approximation ratio for \sst in the special case of submodular costs (assuming ETH).

We show Theorem~\ref{thm:submod-hardness} via a reduction from the densest-$k$-subgraph problem.

The setting for the hard instance consists of $n$ tests, each of which requires multiple machines to be activated.
Formally, there are $m$ machines and each test $i\in [n]$ is associated with a subset $M_i\subseteq [m]$ 
 of machines that need to be activated.
For any batch of tests $B\subseteq [n]$, the cost $c(B) = \abs{\bigcup_{i\in B} M_i}$ is the number of machines required to conduct every test in $B$.
Observe that $c$ is a coverage function: so it is monotone and submodular.

One might notice a similarity between this setting and the one in \S\ref{sec:or_testing}.
The setting in \S\ref{sec:or_testing} has tests that just require \textit{any one} of the machines in $M_j$ to be activated (OR-condition), while this new setting requires \textit{all} machines $M_j$ to be activated (AND-condition).
This subtle distinction produces vastly differing hardness results.

\begin{figure}
     \centering
     \begin{subfigure}[b]{0.4\textwidth}
         \centering
         \includegraphics[width=\textwidth]{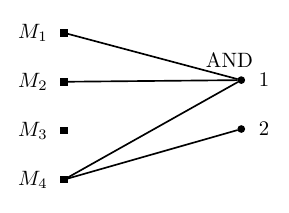}
         \caption{Conducting tests $\{1, 2\}$ requires activating machines $\{M_1, M_2, M_4\}$.}
         \label{fig:y equals x}
     \end{subfigure}
     \hfill
     \begin{subfigure}[b]{0.4\textwidth}
         \centering
         \includegraphics[width=\textwidth, page=2]{figures/matching_machines.pdf}
         \caption{Conducting tests $\{1, 2\}$ only requires activating machines $\{M_4\}$.}
         \label{fig:three sin x}
     \end{subfigure}
     \hfill
        \caption{These special cases of subadditive cost function differ by whether a test requires the AND of all machines or the OR of all machines.}
        \label{fig:matching-machines}
\end{figure}

We reduce from the densest $k$-subgraph (\dks) problem, which is defined as follows.
An instance of \dks consists of a graph $G = (V, E)$, and an integer $k\in \mathbb{Z}_+$.
For a given set $S\subseteq V$, we say that an edge $(u, v)$ is induced by $S$ if both $u\in S$ and $v \in S$.
The goal of \dks is to find subset $S\sse V$ of size $k$ such that the number of induced edges is maximized: 
$$\max_{S\sse V : \abs{S}=k} \,\,\abs{(u, v)\in E\,:\, u, v \in S}.$$

Our reduction uses the minimization version of \dks, known as the \textit{densest $r$ edges} (\dre) problem.
An instance consists of a graph $G = (V, E)$ and a target $r\in \mathbb{Z}_+$.
The objective is to find set $A\sse V$ of minimum cardinality such that there are $r$ induced edges,
i.e.
$$\min \left\{\abs{A} \,\,: \,\, A \subseteq V, \, \abs{(u, v) \in E : u \in A, v\in A} \ge r \right\}.$$

We crucially use the following hardness result.
\begin{theorem}[\cite{Manurangsi17}]\label{thm:dks-hardness}
    \sloppy Assuming the exponential time hypothesis, there is no $n^{1/\poly(\log \log n)}$-approximation to $\dks$ or \dre.
\end{theorem}
While the result in \cite{Manurangsi17} is stated for \dks, it is well known that the approximation ratios for \dks and \dre are polynomially related, i.e., an $\alpha$-approximation for one implies an $O(\alpha^2)$-approximation for the other.
We will also work with a bicriteria approximation algorithm for $\dre$: 

\begin{definition} \label{def:bicrit-apx}
    An $(\alpha, \beta)$-bicriteria approximation algorithm for \dre finds a solution with at most $\alpha\cdot \opt$ nodes and at least $r / \beta$ edges, where $\opt$ is the optimal value of the \dre instance.
\end{definition}
Using a standard set-cover argument, we obtain the following:
\begin{lemma}\label{lem:bicrit-approx-conversion}
    If there is an $(\alpha, \beta)$-bicriteria approximation to \dre there there is an $O(\alpha\beta \cdot \log r)$ approximation to \dre.
\end{lemma}
\begin{proof}
We maintain a solution $S\sse V$ which is initially empty.
Let $E(S)$ denote the number of induced edges in $S$.
As long as $E(S)<r$, we do the following.
(i) Apply the $(\alpha, \beta)$-bicritera approximation for \dre on the subgraph $G[V\setminus S]$ with target $r'=r-E(S)$.
(ii) Let  $S'\sse V\setminus S$ be the solution obtained.
(iii) Update the solution $S\gets S\cup S'$.
Note that the optimal value of each \dre instance is at most $\opt$ (the original optimal value) because we reduce the target $r'$ appropriately.
By the $(\alpha,\beta)$ bicriteria guarantee, the number of new edges added in each iteration is at least $r'/\beta$ and the number of new nodes is at most $\alpha \opt$.
By a set-cover type analysis, the number of iterations is at most $\beta \ln r$.
So the number of nodes in the final solution is at most $(\alpha \beta \ln r) \cdot \opt$.
\end{proof}

We are now ready to relate the submodular-cost \sst problem to \dre.
\begin{lemma} \label{lem:bicriteria-apx}
    If there is an $\alpha$-approximation to $\sst$ with submodular costs then there is a $(4 \alpha, 2\ln \abs{E})$-bicriteria approximation to \dre.
\end{lemma}
\begin{proof}
Given an instance of \dre, with $G = (V, E)$ and $r$ the instance of \sst is as follows.
Each edge $e \in E$ is a test, and each node $v \in V$ is a machine.
Performing any test $e = (u, v)$ requires both machines $u$ and $v$ to be activated: so $M_e=\{u,v\}$.
Furthermore, each test fails with identical probability $q=\frac{\ln \abs{E} }r$.
We assume, without loss of generality, that $r>\ln \abs{E}$: as \dre has a trivial logarithmic approximation when $r\le \ln \abs{E}$.

The first step is to establish 
\begin{equation}\label{eqn:lb-dre}
    \opt_{\sst} \leq 2 \opt_{\dre}.
\end{equation}
Let $k = \opt_{\dre}$, and let $S^*$ be the optimal set picked by \dre.
A feasible solution to \sst is to test edges in $S^*$ in the first batch followed by testing the remaining $E \setminus S^*$ edges in the second batch.
Clearly, the set $S^*$ contains at least $r$ tests (edges) and requires activating $k$ machines (nodes).
The remaining $\abs{E}-k$ tests are conducted only when all tests in the first batch pass, which occurs with probability at most $(1-q)^r$.
Thus, the expected cost of this \sst solution is at most
$$k + (1-q)^r\cdot \abs{E} = k + \left(1- \frac{\ln \abs{E} }{r} \right)^r \abs{E} \,\,\leq\,\, k + 1 \leq 2k,$$
which proves \eqref{eqn:lb-dre}.

Now, given any solution to \sst with cost $\alg_\sst$, we show how we can recover a solution $S\sse V$ to \dre such that:
\begin{equation} \label{eqn:ub-dre}
    \abs{S} \leq 2 \alg_\sst \quad \mbox{ and }\quad E(S)\ge \frac{r}{2\ln \abs{E}}.
\end{equation}

Fix any solution $\batch=\langle B_1, \dots, B_\ell \rangle$ to $\sst$.
Let $j$ be the largest index such that $\prod_{i = 1}^{j-1} P(B_i) \geq \frac{1}{2}$.
We recover a solution for $\dre$ by picking the edges $B_1 \cup B_2 \cup \dots \cup B_j$ and all the nodes $S$ contained in these edges.
Now,
$$\alg_\sst =\ECost{\batch} \geq \frac{1}{2}\sum_{i = 1}^j c(B_i) \geq \frac{1}{2} c(\bigcup_{i=1}^j B_i) = \frac{1}{2} \abs{S}.$$

It remains to show that the number of edges $E(S)$ is large.
Note that $E(S)\ge \sum_{i=1}^j \abs{B_i}$ by definition of $S$.
Moreover, by choice of index $j$, 
$$\frac{1}{2} > \prod_{i = 1}^{j} P(B_i) = (1-q)^{\sum_{i=1}^j \abs{B_i}} \ge (1-q)^{E(S)} = \left(1-\frac{\ln \abs{E}}{r}\right)^{E(S)}.$$
It now follows that $E(S)\ge \frac{\ln 2}{\ln \abs{E}}r$.
This completes the proof of \eqref{eqn:ub-dre}.

If we use an $\alpha$-approximation algorithm for \sst with submodular cost, we obtain $\alg_{\sst} \leq \alpha\opt_{\sst}\le 2\alpha \opt_{\dre}$ by \eqref{eqn:lb-dre}.
Furthermore, by \eqref{eqn:ub-dre} solution $S$ has at most $2\alg_{\sst} \leq 4\alpha\opt_{\dre}$ nodes and at least $\frac{r}{2\ln \abs{E}}$ edges.
Hence, $S$ is a $(4\alpha, 2\ln\abs{E})$ bicriteria approximation for \dre.
\end{proof}

\sloppy Combining Lemma~\ref{lem:bicrit-approx-conversion} and Lemma~\ref{lem:bicriteria-apx}, we see that any $n^{1/\text{poly}(\log \log n)}$-approximation to submodular-cost \sst contradicts Theorem~\ref{thm:dks-hardness}.
This proves Theorem~\ref{thm:submod-hardness}.

\subsection{NP-Hardness for \sst with  Batch Costs}
Here, we show that the sequential testing problem is NP-hard even for the simple batch-cost structure.
Recall that the batch-cost structure is a special case of subadditive costs, where any subset is an allowed batch with cost equal to  a fixed setup cost \(\beta\) plus the individual  testing costs. Formally, given individual costs $\{c_i\}_{i=1}^n$ and $\beta$, the cost function is 
\[c(B) = \beta + \sum_{i \in B} c_i,\qquad \forall B\sse [n].\]

We refer to this special case as the sequential batch-testing problem (\sbt). We note that this problem was studied in \cite{daldal2016approximation,SegevS22,tanGeneralFrameworkSequential2024} and a PTAS was known. However, the computational complexity of the problem was open. Here, we show:

\begin{theorem}\label{thm:sbt-hardness}
The     sequential batch-testing problem for series systems (\sbt)   is NP-hard.
\end{theorem}

\paragraph{Relating optimal values of \sbt and the ratio problem.} Our starting point is a lower bound on the \sbt optimal value in terms of the    ratio problem. Let  
\[\rho^* = \min_{\emptyset \neq S \sse [n]} \frac{c(S)}{1-P(S)}\]
denote the optimal ratio of any batch. 
Recall that for any subset \(S \sse [n]\), \(P(S) = \prod_{i\in S} p_i\) is the probability that every test in the subset passes. 
\begin{lemma}\label{lem:ratio-lb}
    Consider any solution \(\batch = \langle B_1, B_2, \dots, B_k\rangle\) to \sbt. Then
    \[\cost(\batch) \ge \rho^*\cdot \left(1-P([n])\right).\]
    Moreover, we have equality if and only if \(c(B_j) = \rho^*\left(1-P(B_j)\right)\) for every batch \(j \in [k]\).
\end{lemma}
\begin{proof}
    Let \(\batch = \langle B_1, B_2, \dots, B_k\rangle\) be any solution to \sbt; note that $\batch$ forms a partition of $[n]$. We use the shorthand \(P_{\le j} = \prod_{\ell=1}^{j} P(B_\ell)\) for \(j \in \{0, 1, \dots, k\}\); note  that \(P_{\le 0} = 1\).    We have
    \begin{align}
    \cost(\batch) &= \sum_{j=1}^k P_{\le j-1} \cdot c(B_j) \,\,\ge\,\, \rho^* \sum_{j=1}^k P_{\le j-1} \cdot \left(1-P(B_j)\right) \label{eq:np-hard-1}\\
    &= \rho^* \sum_{j=1}^k \left(P_{\le j-1} - P_{\le j}\right) \,\,=\,\,  \rho^* \left(1 - P_{\le k}\right) \,\,=\,\, \rho^* \left(1-P([n])\right).\notag
\end{align}
Above, the inequality uses the fact that for all \(j \in [k]\),
\[c(B_j) = \frac{c(B_j)}{1-P(B_j)} \left(1-P(B_j)\right) \ge \rho^* \left(1-P(B_j)\right).\]
This proves the first statement. Moreover,  equality holds in \eqref{eq:np-hard-1}
if and only if \(c(B_j) = \rho^*\left(1-P(B_j)\right)\) for all \(j \in [k]\). This proves the second statement in the lemma.
\end{proof}

\paragraph{Reduction from \(3\)-Partition.}
Our proof relies on a reduction from \(3\)-Partition~\cite{GareyJ79}. Recall that a \(3\)-Partition instance consists of \(3m\) integers \(a_1, a_2, \dots, a_{3m}\) and an integer \(b\), with \(\sum_{i=1}^{3m} a_i = mb\) and each \(a_i\) satisfying \(b/4 < a_i < b/2\). The goal is to partition the $3m$ integers into \(m\) triplets, such that the sum of elements in each triplet equals \(b\). The decision problem outputs true if such a partition exists and false otherwise. This problem is {\em strongly} NP-hard, i.e., assuming $\text{P}\ne\text{NP}$, there is no algorithm for \(3\)-Partition running in time polynomial in $m$ and $b$. 

Given any instance of \(3\)-Partition, we create an instance of \sbt with \(n = 3m\) tests as follows.
\begin{itemize}
    \item The cost of test \(i \in [n]\) is \(c_i = a_i\).
    \item The probability of test \(i \in [n]\) passing is \(p_i = 2^{-a_i}\).
    \item The setup cost is \(\beta = R - b - 3/2\), where \(R = 3\cdot 2^{b-1}\).
    \item The target is \(T = R\left(1-P([n])\right) = R\left(1-2^{-mb}\right)\). 
\end{itemize}
The goal in the \sbt instance is to determine if there  is a solution of expected cost at most $T$. 

Note that all the costs and probabilities in the \sbt instance are rational, where the largest integer in their rational representation is \(O(2^b)\), hence the instance size is \(\poly(m, b)\). We note that this is indeed larger than the \(3\)-Partition instance size, which is \(\poly(m, \log b)\).  However, \(3\)-Partition is strongly NP-hard~\cite{GareyJ79}, so 
this reduction suffices to prove Theorem~\ref{thm:sbt-hardness}. We note that this establishes ``weak'' NP-hardness of \sbt.

For simplicity, for any subset \(S\sse [n]\), we define
\(a(S) \coloneqq \sum_{i \in S} a_i\)
so that
$$ c(S) = \beta + a(S) = R-b-\frac32 +a(S) \quad \mbox{ and } \quad 1-P(S) = 1-2^{-a(S)}.$$  
The following lemma captures the key property of our   instance.
\begin{lemma}\label{lem:sbt-ratio}
    For any non-empty subset of tests \(S\sse [n]\), we have
    \[c(S) \ge R\cdot \left(1-P(S)\right),\]
    with equality if and only if \(a(S) = b\).
\end{lemma}
\begin{proof}
    Fix any non-empty subset \(S\sse [n]\). We substitute the definitions of \(c(S)\) and \(P(S)\) to obtain
    \begin{equation} \label{eq:diff-def}
        c(S) - R\left(1-P(S)\right) \,\,=\,\, R \cdot 2^{-a(S)} + a(S) - b - \frac{3}{2} \,\, =\,\, g(a(S)),
    \end{equation}
    where we define  function \(g(x) := R \cdot 2^{-x} + x - b - \frac{3}{2}\). 
    It is easy to see that $g(x)$  is convex and satisfies \(g(b) = 0\). Moreover, \(g(b-1) = 1/2\) and \(g(b+1) = 1/4\), so by convexity \(g(x) \ge 1/4\) for all \(x \in \mathbb{Z}\setminus \{b\}\).
Thus, 
$$ c(S) - R\left(1-P(S)\right) \,\, =\,\, g(a(S)) \,\, \ge\,\, 0, \quad \forall S\sse[n]. $$
Moreover, we have equality above if and only if $a(S)=b$. 
\end{proof}
Using Lemma~\ref{lem:sbt-ratio} it follows that the optimal ratio \(\rho^* \ge R\). Combined with Lemma~\ref{lem:ratio-lb}, the optimal \sbt value $\opt\ge R(1-P([n])) = T$ (irrespective of whether the \(3\)-Partition instance has a solution). We now prove that $\opt=T$ if and only if \(3\)-Partition   has a solution. 

\begin{itemize}
    \item Suppose that \(3\)-Partition has a solution, i.e.,  we have a partition  \(S_1, S_2, \dots, S_m\) of $[n]$ such that \(a(S_i) = b\) for all \(i \in [m]\). Consider $\batch = \langle S_1, S_2, \dots, S_m\rangle$ as a solution to \sbt. By Lemma~\ref{lem:sbt-ratio}, 
\(c(S_i) = R\left(1-P(S_i)\right)\) for every \(i \in [m]\), which also implies that \(\rho^* = R\). Since \(c(S_i) = \rho^*\left(1-P(S_i)\right)\) for every \(i\in[m]\), Lemma~\ref{lem:ratio-lb} gives
\[\opt \le \cost( {\batch}) = \rho^*\left(1-P([n])\right) = R\left(1-P([n])\right) = T.\]

    \item Conversely, suppose that $\opt\le T$. Let \(\batch = \langle B_1, B_2, \dots, B_k\rangle\) such that \(\cost(\batch) \le T\). So, 
    \[T \ge \cost(\batch) \ge \rho^*\left(1-P([n])\right) \ge R\left(1-P([n])\right) = T.\]
The second inequality is by   Lemma~\ref{lem:ratio-lb} and the third inequality uses  \(\rho^* \ge R\). So, all the inequalities above must be equalities, i.e.,  \(\rho^* = R\) and $\cost(\batch) = \rho^*\left(1-P([n])\right)$. Applying the equality case of Lemma~\ref{lem:ratio-lb} now gives us
\(c(B_j) = R\left(1-P(B_j)\right)\) for all \(j \in [k]\). Using Lemma~\ref{lem:sbt-ratio} for each $B_j$ yields  \(a(B_j) = b\)  for all $j\in[k]$. Finally, as the batches \(B_1, B_2, \dots, B_k\) partition \([n]\), this is a solution to the \(3\)-Partition instance.
\end{itemize}
This completes the proof of Theorem~\ref{thm:sbt-hardness}.

\section*{AI Disclosure}
The NP-hardness proof of sequential batch testing (Theorem~\ref{thm:sbt-hardness}) was obtained by Claude Opus 5.5: we  gave a prompt to prove NP-hardness, together with the previous preprint of our paper.
The authors checked and rewrote the proof with a focus on exposition, and take full responsibility for its correctness.

\bibliographystyle{alpha}
\bibliography{references}

\newpage
\appendix

\section{Proof of approximation of MSSC} \label{app:mssc-proof}
Recall that an instance of \msc consists of a set $E$ of elements with weights $\{w_e\}_{e\in E}$ and $M$ subsets $\{S_i\sse E\}_{i=1}^M$ with costs $\{c_i\}_{i=1}^M$.
An \msc solution is a permutation $\sigma=\langle \sigma(1), \sigma(2),\dots, \sigma(M) \rangle$ of the $M$ sets.
Given solution $\sigma$, the {\em cover-time} of any element $e\in E$, denoted $\cov(\sigma,e)$, is the cost of the smallest prefix of $\sigma$ that covers $e$.
That is, if $e\in S_{\sigma(j)} \setminus \left(S_{\sigma(1)} \cup \dots \cup S_{\sigma(j-1)} \right)$ then $\cov(\sigma,e) = c_{\sigma(1)} + \dots+ c_{\sigma(j)}.$
The objective in \msc is to minimize the total weighted cover time $$\sum_{e\in E} w_e\cdot \cov(\sigma,e).$$

The greedy algorithm for \msc works as follows.
If $R$ denotes the set of uncovered elements (initially $R=E$) then we select the set $S_i$ that minimizes the score $$\score(S_i) = \frac{c_i}{\sum_{e\in S_i\cap R} w_e}.$$

We consider the situation where we cannot solve the greedy choice problem optimally, but can only get a $\rho$-approximation,
i.e., we can obtain set $\hat{S}$ such that
$$\score(\hat{S})\leq \rho\cdot \min_{i} \score(S_i).$$

The original proof of \cite{FeigeLT04} assumed that all costs are uniform ($c_i = 1$) and that the greedy choice can be solved exactly. 
\cite{GolovinGKT08} adapted this proof for non-uniform costs (and even more general $L_p$ norms), still assuming the greedy choice can be computed exactly. Here, we adapt the proof from \cite{FeigeLT04} to show how a $4\rho$-approximation can be obtained for non-uniform costs assuming a $\rho$-approximation for the greedy choice problem.

\begin{theorem}
    There is a $4\rho$-approximation to the MSSC problem, where $\rho$ is the approximation ratio of the greedy choice problem.
\end{theorem}

\begin{proof}
The proof of \cite{FeigeLT04} relies on plotting a curve for $\opt$ and $\grd$ each, where the area under each curve is the value of $\opt$ and $\grd$ respectively. We let $O$ be the $\opt$ curve and $G$ be the $\grd$ curve (both are defined shortly).
The goal is to show that shrinking the area of $G$ by a factor of $4\rho$ allows it to fit entirely within $O$, which implies $\grd \leq 4\rho\cdot \opt$ as desired.

\paragraph{OPT curve.} The optimum curve $O$ consists of $\abs{E}$ columns corresponding to the elements. The column for any element  $e\in E$ has width $w_e$ and height $\cov(\opt, e)$. 
 The columns in $O$ are sorted by increasing height.
This yields a monotone increasing curve. Clearly, the area under curve $O$ equals $\opt$.

\paragraph{GRD curve.} 
For any step $i$ in the greedy solution, define the following:
\begin{itemize}
    \item $X_i$ is the set of elements covered in step $i$.
    \item $R_i = E - \bigcup_{j=1}^{i-1}X_i$ is the set of uncovered elements at the start of step $i$. 
    \item $s_i$ is  the  cumulative cost at step $i$. That is,   $s_0 = 0$ and $s_i = s_{i-1} + c_i$ for $i\ge 1$. 
    \item $P_i = \frac{c_i\cdot w(R_i)}{w(X_i)}$ is the {\em price} at  step $i$.
\end{itemize}
The greedy curve $G$ also consists of $\abs{E}$ columns corresponding to elements. The column for any element  $e\in E$ has width $w_e$ (as for $O$), but the   height is set to   $p_e = P_i$ if $e \in X_i$. The elements are ordered by their cover-time in $\grd$: so   elements covered first are placed towards the left.   It can be seen that the area under $G$ gives the value of $\grd$, since
$$\text{Area} = \sum_e w_e\cdot p_e
 =\sum_i w(X_i)\cdot  P_i = \sum_i c_i w(R_i) = \sum_i (s_i-s_{i-1}) w(R_i)=\sum_i s_i w(X_i) = \grd,$$
where the second last inequality follows from the fact that $c_i = s_i - s_{i-1}$ and $w(X_i) = w(R_i) - w(R_{i+1})$.
Note that the greedy curve $G$  is not  monotonically increasing.

Let $G'$ be the scaled variant of $G$, where $G$ is scaled down by $2$ along the horizontal axis, and by $2\rho$ along the vertical axis.
To show that $G'$ fits within $O$, we align $G'$ rightwards such that the bottom right corner of $G'$ aligns with the bottom right corner of $O$. 

We then show that any arbitrary point $q'$ in $G'$ is also within $O$.
Let $q=(x,y)$ be the point in  the original greedy curve $G$ that corresponds to $q'$.
Point $q$ corresponds to some element $e\in E$, which is covered at some step $i$ in greedy (i.e. $e\in X_i$).
This implies that its height $y\leq \frac{c_i w(R_i)}{w(X_i)}$. Moreover, $q$ is at most $w(R_i)$ distance away from the right boundary of  $G$. 
After scaling, the height $h$ of  point $q'\in G'$ is at most $\frac{c_i w(R_i)}{2\rho\cdot  w(X_i)}$ and its distance to the right is $r\le \frac{w(R_i)}{2}$.
To  show  that $q'$ lies within $O$ it suffices to show  that the total weight of elements with height at least $h$ in the optimal curve  $O$ is at   least $r$. We will show the following stronger claim: the total weight of elements from $R_i$ that have height (i.e., cover-time) at least $h$ in the optimal curve is at least $w(R_i)/2$. Suppose not: then,  the total weight of elements in $R_i$ that are covered in $\opt$ by time $h$ is more than $w(R_i)/2$. Let $Q\sse [M]$ denote    the sets in $\opt$ with cumulative cost less than $h$: so $\sum_{j\in Q} c_j\le h$. The elements of $R_i$ that are covered in $\opt$ by time $h$ are $\cup_{j\in Q} (S_j\cap R_i)$: so we have $\sum_{j\in Q} w(S_j\cap R_i)> \frac{w(R_i)}{2}$. So,
$$\min_{j\in Q} \frac{c_j}{w(S_j\cap R_i)} \leq \frac{\sum_{j\in Q}c_j}{\sum_{j\in Q}w(S_j \cap R_i)} < \frac{h}{w(R_i)/ 2 }  \le  \frac{c_i}{\rho\cdot w(X_i)}.$$
This implies that the best  greedy choice in step $i$ has ratio less than  $\frac{c_i}{\rho\cdot w(X_i)}$. As our algorithm uses a $\rho$-approximate greedy choice, we must have $\frac{c_i}{w(X_i)} < \rho\cdot \frac{c_i}{\rho\cdot w(X_i)}$, which is a contradiction. 
\end{proof}

\end{document}